\documentclass[aps,pra,twocolumn,amsmath,amssymb,showpacs]{revtex4-1}

\usepackage{graphicx}
\usepackage[ansinew]{inputenc}
\usepackage{ae,aecompl}
\usepackage{textcomp}
\usepackage{hyperref}

\begin{document}

\title{Generation of single photons from an atom-cavity system}

\author{Martin~Mücke}
\altaffiliation[Present address: ]{Kayser-Threde GmbH, Perchtingerstrasse 5, 81379 Munich, Germany.}
\author{Joerg~Bochmann}
\altaffiliation[Present address: ]{Department of Physics, and California NanoSystems Institute, University of California, Santa Barbara, CA 93106, USA.}
\author{Carolin~Hahn}
\author{Andreas~Neuzner}
\author{Christian~Nölleke}
\author{Andreas~Reiserer}
\author{Gerhard~Rempe}
\author{Stephan~Ritter}
\email{stephan.ritter@mpq.mpg.de}
\affiliation{Max-Planck-Institut für Quantenoptik, Hans-Kopfermann-Strasse 1, 85748 Garching, Germany}

\begin{abstract}
A single rubidium atom trapped within a high-finesse optical cavity is an efficient source of single photons. We theoretically and experimentally study single-photon generation using a vacuum stimulated Raman adiabatic passage. We experimentally achieve photon generation efficiencies of up to 34\,\% and 56\,\% on the D$_1$ and D$_2$ line, respectively. Output coupling with 89\,\% results in record-high efficiencies for single photons in one spatiotemporally well-defined propagating mode. We demonstrate that the observed generation efficiencies are constant in a wide range of applied pump laser powers and virtual level detunings. This allows for independent control over the frequency and wave packet envelope of the photons without loss in efficiency. In combination with the long trapping time of the atom in the cavity, our system constitutes a significant advancement toward an on-demand, highly efficient single-photon source for quantum information processing tasks.
\end{abstract}


\maketitle

\section{Introduction}
Single photons as carriers of quantum information are at the heart of many quantum information processing protocols. A prime example is  the proposal of Knill, Laflamme and Milburn \cite{knill2001, kok2007} which relies on deterministic single photon sources and linear optical elements for the realization of conditional quantum gates. Furthermore, single photons are also very well suited for the distribution of information between distant nodes in a quantum network \cite{kimble2008, ritter2012} because of their weak interaction with the environment.

The generation of single photons has been studied in a large variety of physical systems \cite{eisaman2011}. Parametric down-conversion (PDC), for example, is a workhorse in the optics community, but sources of pure single photons  based on PDC suffer from a fundamental efficiency limit of 25\,\% \cite{christ2012}. In this respect, single emitters are not only a natural choice but also offer great promise. Prime examples are single trapped atoms \cite{darquie2005, hofmann2012}, ions \cite{maunz2007, gerber2009}, single molecules \cite{brunel1999, lounis2000, lee2004, trebbia2009, lettow2010} and solid-state-based systems such as quantum dots \cite{santori2001, patel2010, flagg2010, matthiesen2012} or color centers in diamond \cite{gaebel2004, wu2007, bernien2012}. However, only a small fraction of the emitted photons can be collected even with high numerical aperture lenses. This limitation can be overcome by placing the single emitter in a high-finesse optical cavity \cite{kuhn2002, keller2004, mckeever2004, barros2009, michler2000, santori2002, chang2006, press2007, dousse2010, nisbet-jones2011, gazzano2013}. On the one hand, a cavity enhances the single-photon emission into the cavity mode via the Purcell effect \cite{purcell1946}. On the other, the emitted photon travels in a well-defined spatial mode, such that it can be efficiently coupled into a single-mode optical fiber for long-distance communication.

The most direct way for single-photon generation in an atom-cavity system is based on excitation of the system with a laser pulse much shorter than the excited-state lifetime, followed by the Purcell enhanced emission into the cavity \cite{purcell1946, bochmann2008}. In this case, the envelope of the photonic wave packet is fixed, with its length set by the cavity decay time. In contrast, a vacuum-stimulated Raman adiabatic passage (vSTIRAP) \cite{law1997, kuhn1999, kuhn2002, hennrich2003} allows the frequency and wave packet shape of the photon to be controlled over a wide range. The dynamics and the efficiency of the single-photon generation process are governed by an interplay of cavity mode volume, transition dipole matrix elements, temporal amplitude of the pump pulse, and frequency detunings.

Here, we experimentally and theoretically study the single-photon emission process from a single $^{87}$Rb atom on the D$_1$ and D$_2$ lines at 795\,nm and 780\,nm, respectively. The main focus is on the achievable efficiencies while tuning the frequency and wave packet envelope of the photon. We concentrate on the photon emission on the respective $F=1\leftrightarrow F'=1$ transition because it has proven very useful for the generation of entangled photon states \cite{wilk2007, weber2009} and the establishment of remote matter-matter entanglement \cite{ritter2012}. The particular interest in the tunability is motivated by the idea of hybrid quantum network architectures in which different atomic systems are connected via a photonic channel \cite{lettner2011}. We find that the vSTIRAP scheme works reliably and with high efficiency over a large parameter range. We want to point out that the reflectivity of our cavity mirrors is not identical, but a designated outcoupling mirror is employed such that an intracavity photon is emitted into a single propagating output mode with 89\,\% efficiency. This high directionality and single-mode character of photon emission are of the utmost importance for the usefulness of any single-photon source.

\begin{figure}
\includegraphics[width=1.0\columnwidth]{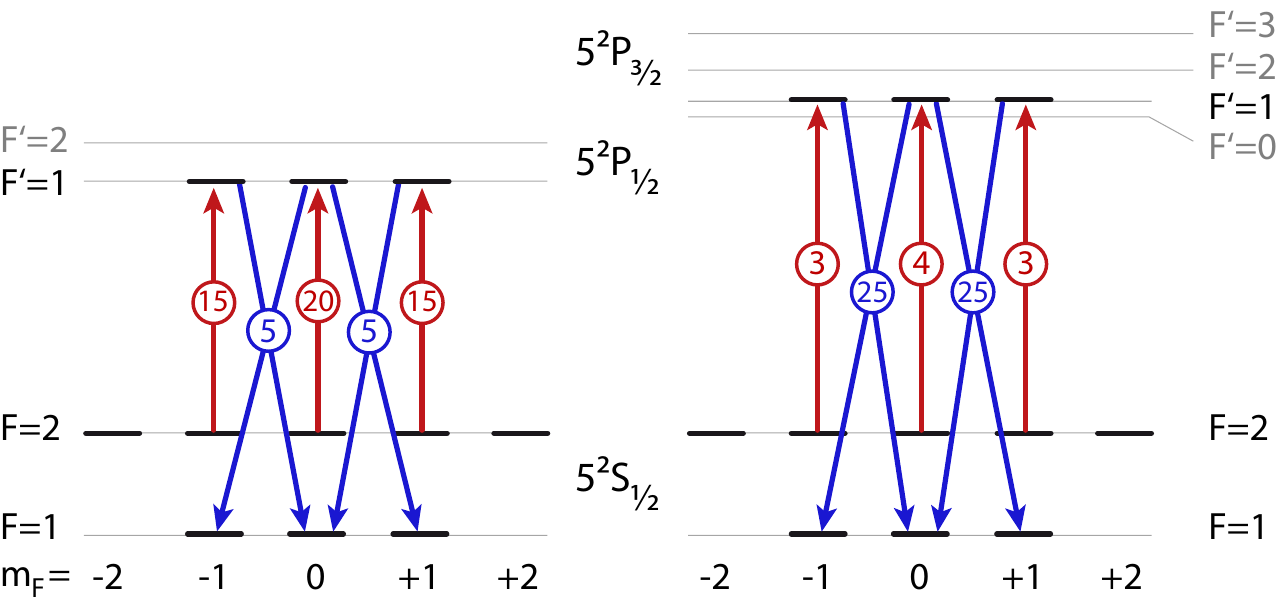}
\caption{\label{fig:scheme}
(Color online) Level scheme of $^{87}$Rb for the D$_1$ (left) and D$_2$ (right) lines at 795 and 780\,nm, respectively (not to scale). Shown are the Zeeman substates of the two hyperfine ground states $F=1,2$ and the excited state $F'=1$. In our scheme, the cavity is resonant with the $F=1\leftrightarrow F'=1$ transition, whereas a $\pi$-polarized control laser pulse (resonant with the $F=2\leftrightarrow F'=1$ transition) drives the single-photon generation process. The numbers in circles indicate the relative transition probabilities.}
\end{figure}

In our vSTIRAP scheme (Fig.\,\ref{fig:scheme}), the cavity is resonant with the \mbox{$F=1\leftrightarrow F'=1$} transition and a $\pi$-polarized control laser addresses the \mbox{$F=2\leftrightarrow F'=1$} transition. Here, unprimed and primed labels refer to the $5^{2}S_{1/2}$ ground and the $5^{2}P$ excited states, respectively. With all atomic population ideally initialized in the $|F,m_F\rangle=|2,0\rangle$ hyperfine ground state, a successful population transfer to $F=1$ coincides with the deposition of a single photon into the cavity mode. Simultaneously with the photon generation, the cavity field decay with rate $\kappa$ then results in the emission of the photon into one well-defined spatiotemporal mode.

The atomic level structures of the D$_1$ and D$_2$ lines exhibit different characteristics. In our scheme, the D$_2$ line offers a five times stronger transition probability compared to the D$_1$ line. The atom-cavity coupling is accordingly a factor of $\sqrt{5}$ larger than that on the D$_1$ line, because the coupling between the atomic transition and the cavity mode is proportional to the dipole matrix element of the transition. One can therefore expect that the larger coupling constant $g$ on the D$_2$ line goes along with higher efficiencies for the emission of a single photon into the cavity. The D$_1$ and D$_2$ lines differ further in terms of the number of excited hyperfine levels and their mutual separation in frequency space. While on the D$_1$ line there exist only two excited hyperfine levels separated by 815\,MHz, the D$_2$ line reveals a more complex structure. There are four excited hyperfine levels with frequency splittings ranging from 72 to 267\,MHz. Although this rich level structure might not directly affect the particular single-photon generation process described in this paper, the presence and finite separation of additional excited hyperfine states can influence the fidelity with which entanglement protocols or quantum memory experiments can practically be implemented \cite{weber2009, specht2011, lettner2011, ritter2012}.

\section{Experimental implementation}
\begin{figure}
\includegraphics[width=1.0\columnwidth]{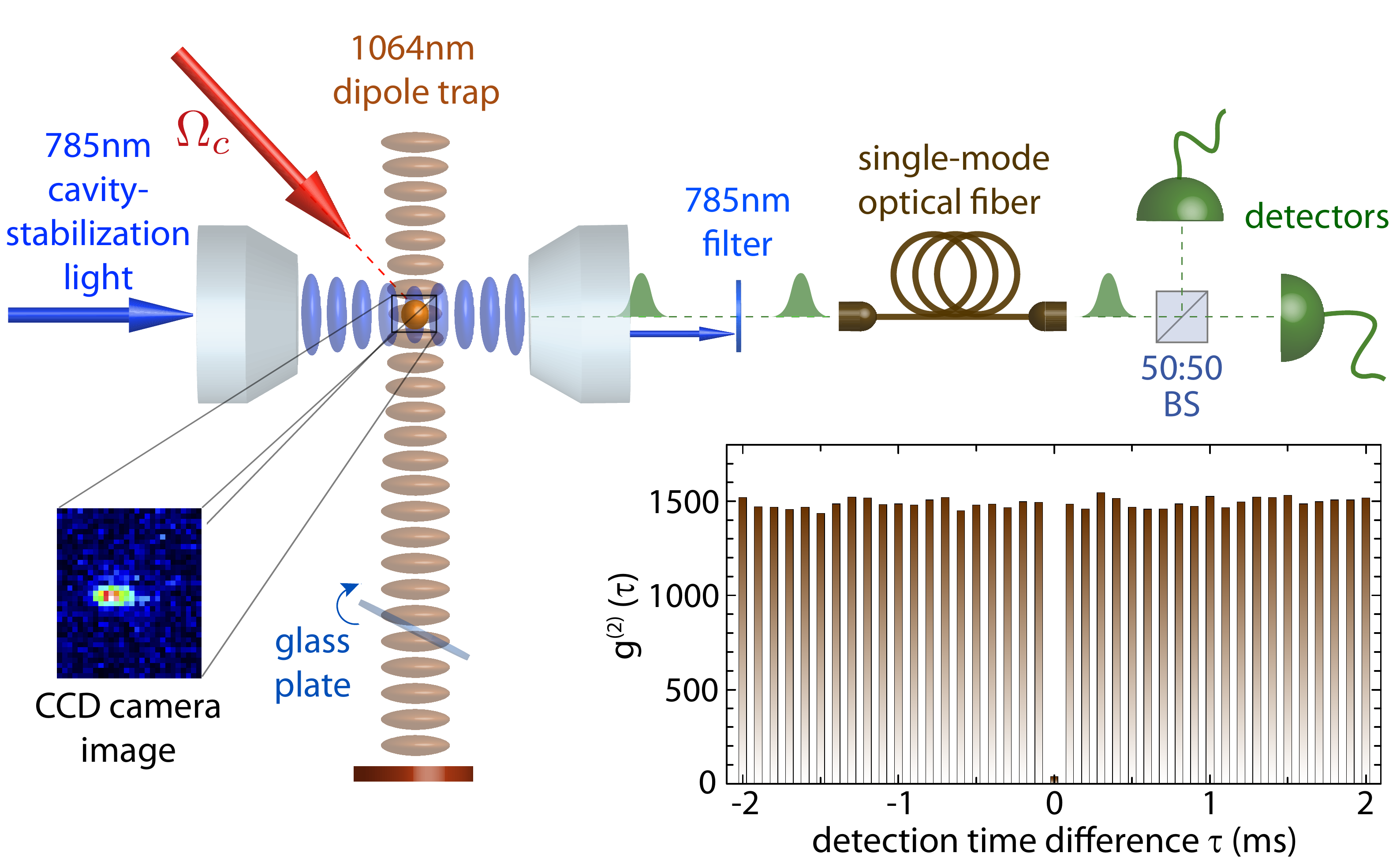}
\caption{\label{fig:setup}
(Color online) Individual $^{87}\mathrm{Rb}$ atoms are trapped in a standing-wave dipole trap at the center of a high-finesse optical cavity. Laser beams perpendicular to the cavity axis are applied for optical cooling, state preparation (not shown), and single-photon generation. Single photons emitted from the cavity are coupled into a single-mode optical fiber and directed to a detection setup. The atom is monitored on a CCD camera by collecting light that is scattered primarily during cooling intervals. A tiltable glass plate allows positioning of the atom in the center of the cavity mode with \textmu m precision. The dimensions of the shown camera image are 15\,\textmu m$\times$19\,\textmu m. Inset: a typical second-order correlation function of the photons emitted from a single atom that was trapped for 30\,s. The detected photons show excellent suppression of coincidence events: $g^{(2)}(0)=2\,\%$, consistent with our background noise.}
\end{figure}
In our experiment (Fig.~\ref{fig:setup}), a single $^{87}$Rb atom is quasipermanently trapped inside a high-finesse optical cavity \cite{bochmann2008}. The maximum atom-cavity coupling constants $g/2\pi$ for the relevant \mbox{$F=1\leftrightarrow F'=1$} transition of the D$_1$ and D$_2$ line are 2.3 and 5.1\,MHz, respectively. With our cavity field and atomic polarization decay rates $\kappa/2\pi = 2.8\,\mathrm{MHz}$ and $\gamma/2\pi=3\,\mathrm{MHz}$ our system operates in the intermediate-coupling regime of cavity QED. The transmission of the cavity mirrors is asymmetric such that photons preferentially exit the resonator through the higher transmission mirror. The output directionality of $(89\pm2)\,\%$ is the ratio of the transmission of the outcoupling mirror and the total round-trip losses including transmission. The latter are inferred from independent measurements of the cavity linewidth and its free spectral range, while the transmission of the outcoupling mirror is determined from measurements of the cavity transmission and reflection \cite{hood2001}. The cavity output mode is coupled efficiently into a single-mode optical fiber and directed to a detection setup consisting of single-photon counting modules (SPCM). The total detection probability for a single photon present in the cavity was $(27.8\pm1.6)\,\%$ and $(16\pm1.6)\,\%$ for the measurements at 780 and 795\,nm, respectively. These values are the product of the output directionality of the cavity, the transmission of the optical path, and the quantum efficiency of the SPCMs. The transmission of the optical path is determined using a probe beam transmitted through the cavity. The SPCMs are calibrated against a laser power meter using optical attenuators of well-known transmission. The uncertainty in the calibration of the employed laser power meter is the main systematic error in our measurements of the photon generation efficiency (relative error of $\pm5$\,\%).

The trapping potential for the single atoms is provided by a horizontal standing-wave dipole trap at 1064\,nm (power 2.5\,W, linear polarization, 3\,mK trap depth deduced from a measured Stark shift on the D$_2$ $|2,0\rangle \leftrightarrow |1,0\rangle$ transition of 110\,MHz). Intracavity light at 785\,nm (linear polarization) is used to stabilize the length of the cavity to be resonant with the \mbox{$F=1\leftrightarrow F'=1$} atomic transition of the D$_1$ or D$_2$ line. The resulting dipole potential (depth $\leq 0.1$\,mK) is much shallower than the trap at 1064\,nm. Typical atom trapping times are tens of seconds. The atoms in the cavity are monitored by collecting fluorescence light (which is primarily emitted during the cooling intervals) with a high numerical aperture (NA) lens system (NA=0.4, spatial resolution 1.3\,\textmu m) and imaging it onto an electron multiplying charge-coupled-device (EMCCD) camera. By tilting a 5-mm-thick glass plate in front of the retroreflecting mirror for the 1064\,nm standing-wave dipole trap, an atom in the trap can be shifted longitudinally such that along this axis it is trapped in the center of the cavity mode \cite{nussmann2005}. Nevertheless, because of its finite temperature, the atom moves considerably along the cavity axis resulting in an effective coupling constant $g_\mathrm{eff}$ averaged over the periodic structure of the cavity mode function. In addition, the atomic motion in the dipole trap potential also leads to a varying Stark shift and therefore alters the resonance frequency of the atomic transitions.

The photon generation scheme is experimentally implemented as follows. Once a single atom is trapped in the cavity, we optically pump it into the \mbox{$|F,m_F\rangle =|2,0\rangle$} Zeeman state with an estimated efficiency of 0.9. We define the quantization axis to coincide with the cavity axis. Next, a single photon is generated by driving the Raman passage via a $\pi$-polarized control laser pulse propagating perpendicular to the cavity axis. The repeated application of this protocol (repetition rate 10\,kHz) with intermittent cooling intervals results in a stream of single photons emitted from the atom-cavity system. The second-order correlation function $g^{(2)}(\tau)$ for the photons produced from one single atom that was trapped in the cavity for 30\,s is shown in the inset of Fig.~\ref{fig:setup}. The detected photons show clear antibunching as coincidence events at $\tau = 0$ are only 2\,\% of those at $\tau \neq 0$. This number is consistent with the combined background and dark count rate of our detectors and therefore proves the perfect single-photon character of our source.

\section{Simulation}
We theoretically study the single-photon generation process for the parameters of our particular atom-cavity system in order to derive ideally expectable generation efficiencies and to identify crucial parameters. The efficiency of the vSTIRAP is directly related to the robustness of the associated dark state which is a coherent superposition of the coupled hyperfine ground states. In the strong-coupling regime of cavity QED, where the cooperativity parameter $C=g^2/(2\kappa \gamma)\gg 1$, the dark state is very robust with negligible admixture of any excited state. In this parameter regime, single-photon generation can be expected with efficiencies close to unity. The situation is different for a cavity system that operates in the intermediate- or even weak-coupling regime, where the emission process is no longer an ideal \mbox{vSTIRAP}. Partial population of the excited states followed by spontaneous decay dramatically alters the dynamics of the emission process. As a consequence, its theoretical description requires all Zeeman substates of the two hyperfine ground states and the excited $F'=1$ hyperfine state to be taken into account. Additional excited hyperfine states besides $F'=1$ are not considered, since these levels play a negligible role for the particular photon generation process described in this paper.

In our simulation, we start with all the population in the \mbox{$|2,0\rangle$} Zeeman state and with $n=0$ photons in the cavity mode. The density matrix $\rho(t)$ of the coupled system can be determined by numerically solving the time-dependent Master equation, where the decay of the intracavity field with rate $\kappa$ and the spontaneous atomic polarization decay from $F'=1$ with rate $\gamma$ are included as Liouvillian terms. The knowledge of $\rho(t)$ allows for the calculation of the photon generation efficiency into the cavity,
\begin{equation}
\eta = 2\kappa \int{\mathrm{Tr}\!\left( a^{\dagger}a\, \rho(t) \right) dt}\,.
\label{eq:efficiency}
\end{equation}
Here, $a^{\dagger}$ ($a$) denotes the creation (annihilation) operator for a photon in the cavity mode.

\begin{figure}
\includegraphics[width=1.0\columnwidth]{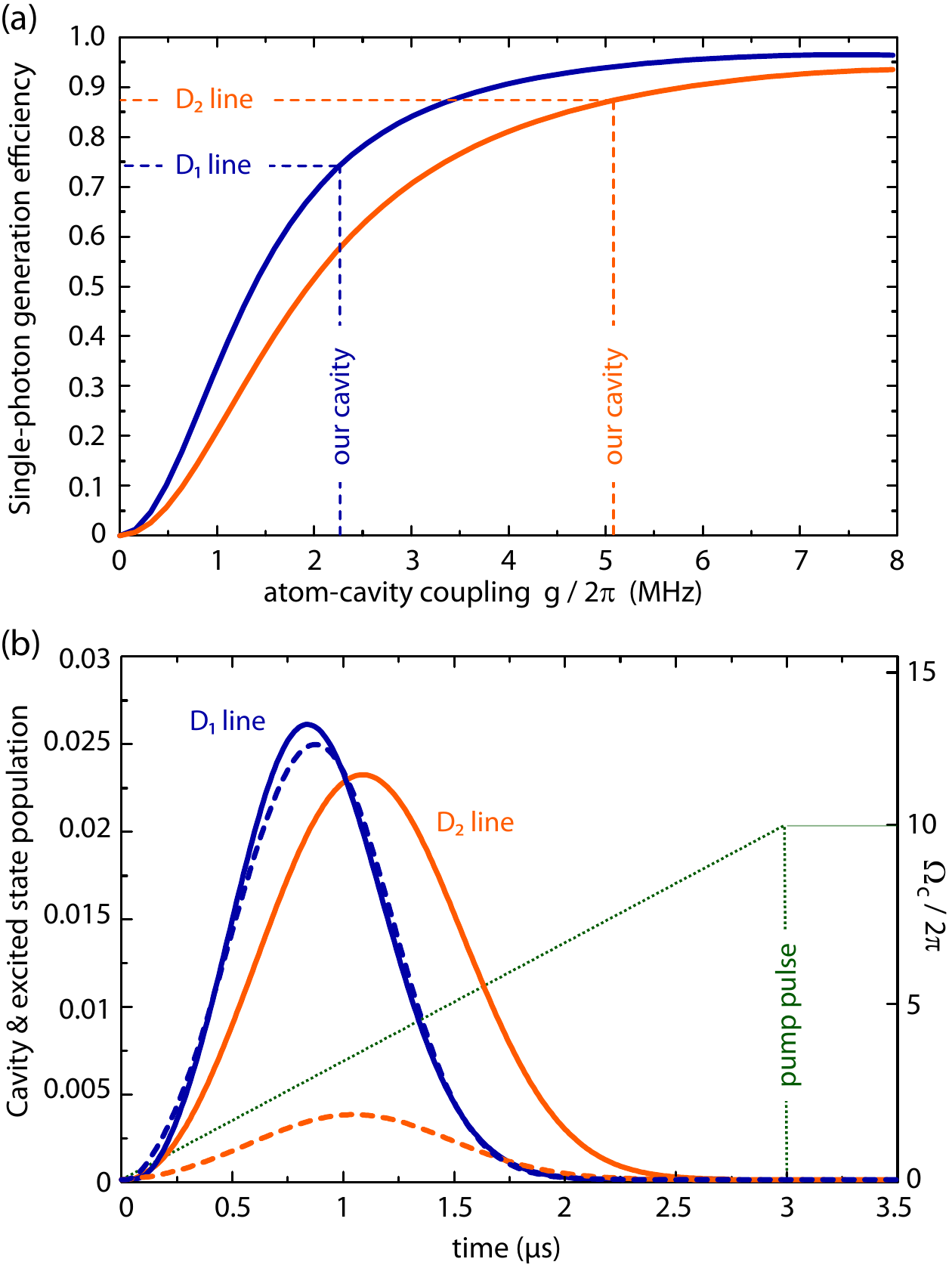}
\caption{\label{fig:theory}
(Color online) Theoretical results. (a) The single-photon generation efficiency into the cavity $\eta$ increases monotonically with $g$, asymptotically approaching unity for strong coupling. The vertical dashed lines represent the parameter set of our cavity setup for atoms maximally coupled to our cavity mode. The expected maximum efficiencies on the D$_1$ and D$_2$ lines are 74\,\% and 87\,\%, respectively.
(b) Corresponding cavity population $n=a^{\dagger}a\, \rho(t)$ (solid lines) representing the single-photon temporal wave packet shape and excited-state population (dashed lines). The green dotted line reflects the profile of the 3\,\textmu-s-long control laser pulse linearly increasing in Rabi frequency.
Parameters: $\kappa=2\pi\times 2.8\,\mathrm{MHz}$, $\gamma=2\pi\times 3.0\,\mathrm{MHz}$.
}
\end{figure}

As a first result, the dependence of the single-photon generation efficiency on the atom-cavity coupling $g$ is shown in Fig.~\ref{fig:theory}(a). In the simulation, the Rabi frequency of the control laser is linearly increased for 3\,\textmu s up to a final value of $\Omega_c^{\mathrm{max}}/2\pi=10\,\mathrm{MHz}$. The final Rabi frequency is chosen such that at the end of the process no population is left in $|2,0\rangle$. The curves for both the D$_1$ and D$_2$ lines increase monotonically and the efficiency approaches unity once the atom-cavity coupling is sufficiently larger than the decay rates $\gamma$ and $\kappa$. For a given atom-cavity coupling, the efficiencies achievable on the D$_1$ line are higher than those on the D$_2$ line. For a given cavity, however, the coupling of a particular atomic transition to the cavity scales with its transition dipole matrix element. The ratio in $g$ for the \mbox{$F=1\leftrightarrow F'=1$} transition on the D$_2$ and D$_1$ lines is $\sqrt{5}$ (see Fig.~\ref{fig:scheme}). For that reason the same cavity gives rise to different couplings for the respective atomic transitions as indicated by the vertical dashed lines. For our cavity, the simulation predicts maximum single-photon generation efficiencies of 74\,\% (D$_1$ line) and 87\,\% (D$_2$ line).

Beyond the absolute efficiencies, the simulation gives further insight into the dynamics of the photon emission process. First of all, the robustness of the dark state depends on the absolute value of the atom-cavity coupling $g$. As a consequence, the larger $g$ is for the respective transition, the less population is transferred to the excited state $F'=1$ during the photon-generation process. This is clearly reflected in Fig.~\ref{fig:theory}(b). Here, the population of the cavity is represented by the solid lines, whereas the dashed lines display the excited-state population. The higher coupling on the D$_2$ line not only delays the emission of the photon from the cavity but also results in less excited-state population than on the D$_1$ line. Excited-state population and successive spontaneous decay paths play a crucial role for the dynamics and---in terms of efficiency---are the main limitations of the studied process. Population in the excited state can decay via emission of a photon into the cavity, which is enhanced via the Purcell effect, or by emission of a photon into free space. However, in our parameter regime only 0.7\,\% (D$_1$ line) and 1.2\,\% (D$_2$ line) of the excited-state population finally result in population of the cavity field and are therefore practically negligible. This argument holds as long as the increase of the control laser pulse intensity is sufficiently slow that the excited state is not immediately populated \cite{bochmann2008}.

The single-photon generation process is finished once the atom is decoupled from the control laser field and the cavity mode. This applies in particular for the atomic hyperfine state $F=1$ and the two Zeeman states $|2,\pm 2\rangle$ (see Fig.\ref{fig:scheme}). The transfer of population into these atomic states without emission of a photon into the cavity is possible via spontaneous decay from the excited state $F'=1$. The respective transition probabilities and branching ratios on the D$_1$ and D$_2$ lines result in a different dynamical behavior and different atomic levels that are preferentially populated. On the D$_2$ line, a branching ratio of 5:1 clearly favors the free-space decay into the $F=1$ hyperfine ground state over the $F=2$ ground state. This process limits the achievable photon generation efficiency. The situation is different on the D$_1$ line where a larger fraction of atomic population is transferred to the excited state during the photon generation process. The excited-state population is comparable to the cavity population and each photon emitted into the cavity goes along with 1.06 photons emitted into free space. With a branching ratio of 1:5, the preferred decay path leads to the $F=2$ hyperfine ground state. The atom is hence reinitialized in one of the $F=2$ Zeeman states and can be addressed again by the control laser. This repeated redistribution of atomic population can transfer the atom into the $|2,\pm 2\rangle$ states, where it is then decoupled from the $\pi$-polarized control laser. On the D$_1$ line, this optical pumping mechanism limits the efficiency of the photon generation process.

\section{Experimental results}
\begin{figure}
\includegraphics[width=1.0\columnwidth]{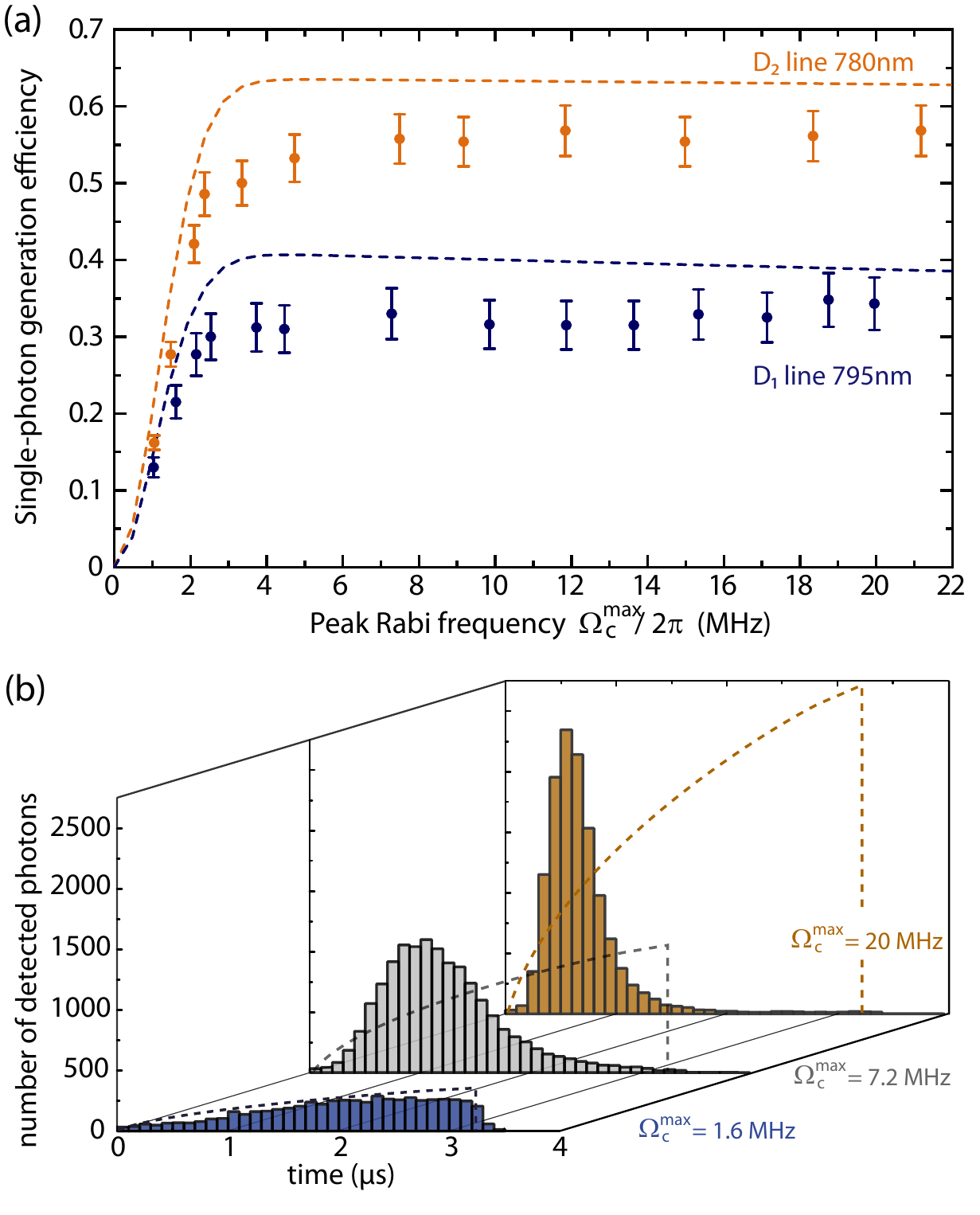}
\caption{\label{fig:data}
(Color online) Photon generation efficiency into the cavity versus maximum control laser Rabi frequency for a 3-\textmu-s long pump pulse.
(a) For low $\Omega_c^{\mathrm{max}}$ the population transfer between the two hyperfine ground states is still incomplete. The generation efficiency therefore rises for increasing peak Rabi frequency until it levels off at the respective maximum observed efficiencies. Error bars are mainly due to the systematic uncertainty of our single-photon detection efficiency, while the statistical error is negligible.
Dashed lines represent the corresponding theory curves for an average coupling constant of $0.5\,g_{\mathrm{max}}$. (b) The histograms display the arrival time distribution of single photons for different control laser pulses (their Rabi frequency is sketched as a dashed line) on the D$_1$ line at 795\,nm. The steepness of the pump pulse directly translates into the length of the emitted photon wave packet.
}
\end{figure}

In a first measurement, we study the influence of maximum control laser Rabi frequency on the single-photon generation efficiency. The applied pump pulse is 3\,\textmu s long and its Rabi frequency has been measured to follow a function $\propto t^{0.75}$ in time up to the maximum value $\Omega_c^\mathrm{max}$. As displayed in Fig.~\ref{fig:data}(a), the measured data for the D$_1$ line at 795\,nm and the D$_2$ line at 780\,nm show similar behavior. In the regime of weak pumping, the efficiency increases with Rabi frequency, as the atomic population transfer from $F=2$ to $F=1$ is still incomplete. At sufficiently high control laser Rabi frequencies, the completion of the population transfer is indicated by a single-photon wave packet that becomes shorter than the length of the control laser pulse [Fig.~\ref{fig:data}(b)]. Here, the generation efficiency saturates at 34\,\% (D$_1$ line) and 56\,\% (D$_2$ line), respectively.

In comparison with the simulation where we assumed maximum atom-cavity coupling, the measured generation efficiencies in our system are consistently lower. As already discussed in Sec.~II, the atomic motion along the cavity axis reduces the coupling constant to an average value of $0.5\,g_\mathrm{max}$ as has been observed in various experimental situations \cite{bochmann2010, muecke2010}. As a consequence, the excited state is significantly populated. The calculated efficiencies for the reduced coupling of $0.5\,g_\mathrm{max}$ are shown as dashed lines in Fig.~\ref{fig:data}(a). To reach the higher values predicted in Sec.~III, the effective coupling needs to be increased to $g_\mathrm{max}$. This can be achieved via better localization of the atom at an antinode of the cavity mode function using stronger confinement along the cavity axis \cite{reiserer2013}.

\begin{figure}
\includegraphics[width=1.0\columnwidth]{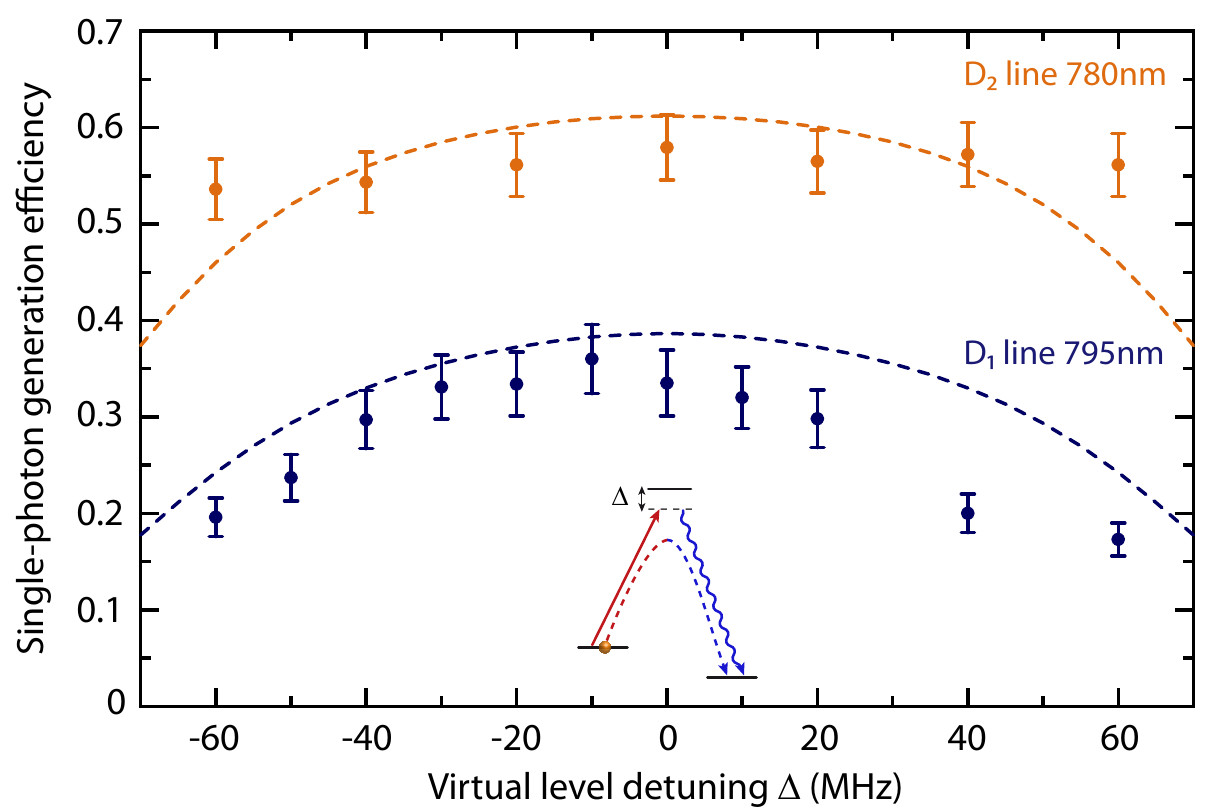}
\caption{\label{fig:data2}
(Color online) Photon generation efficiency into the cavity as a function of virtual level detuning $\Delta$ with respect to the Stark-shifted atomic transition. High and nearly constant efficiency is achieved for the production of single photons tuned over a range of many tens of MHz. Dashed lines represent the theoretical prediction for an assumed atom-cavity coupling $g_\mathrm{eff}=0.5\,g_\mathrm{max}$.}
\end{figure}

The single-photon wave-packet shape can be tailored by different temporal profiles of the pump pulse intensity, as is exemplified in Fig.~\ref{fig:data}(b). The histograms represent the detection time distribution of single photons generated on the D$_1$ line and are therefore the ensemble average of the single-photon wave packet shape. The difference between the three scenarios is the maximum Rabi frequency of the control laser pulse reached after 3\,\textmu s and hence the steepness of the applied pulse (dashed lines). The steeper the slope, the shorter is the temporal extension of the wave packet. We are able to deliver photons with a full width at half maximum ranging from 250\,ns up to several microseconds, while the efficiency of the photon generation process remains unaffected. The vSTIRAP technique allows for the control not only of the length of the photon but also of its overall shape by tailoring the control laser power \cite{keller2004, nisbet-jones2011}.

To generate photons of different frequency, simultaneous tuning of the cavity resonance and the control laser frequency is required. While there is a strong dependence of the single-photon generation efficiency $\eta$ on the two-photon detuning \cite{hennrich2000}, it is very robust with respect to the detuning $\Delta$ from the Stark-shifted atomic transition (Fig.~\ref{fig:data2}). The dashed lines represent the theoretical prediction for an atom-cavity coupling of $0.5\,g_\mathrm{max}$. It is evident that the frequency of the photon is tunable over a wide range of more than 100\,MHz while the generation efficiency remains almost unaffected. The applied control laser pulse is again 3\,\textmu s long with a maximum Rabi frequency $\Omega_c^\mathrm{max}$ of 15\,MHz (D$_1$ line) and 26\,MHz (D$_2$ line), respectively. It has been verified that a decrease of the photon generation efficiency for larger detunings is related to an incomplete population transfer from $F=2$ to $F=1$. Compensating for this with higher $\Omega_c^\mathrm{max}$ should be possible, however it could not be observed with the available control laser powers.

\section{Conclusion}
We have studied the efficiency and dynamics of single-photon emission on the D$_1$ and D$_2$ lines of $^{87}$Rb using a vSTIRAP. The high efficiencies achieved for triggered production of single photons into a single, free-space mode are of great importance for all practical applications of single-photon sources, as single photons generated inside the cavity mode can be coupled into a single-mode optical fiber with an overall efficiency above 0.8. The vSTIRAP technique allows for a broad range of single-photon wave-packet shapes and fine tuning of the photon frequency at nearly constant efficiency. Very recently, we have studied the wave-packet overlap of photons generated from two independent systems using this technique \cite{noelleke2013}. In a Hong-Ou-Mandel-type setup, we find an interference contrast of 64\,\%. We expect this value to increase significantly when fluctuations in the coupling strength and Stark shift are eliminated via better localization of the atom along the cavity axis and cooling to the motional ground state \cite{reiserer2013}. By theoretical modeling, we show that also even higher generation efficiencies are expected with our current cavity setup once these techniques are combined with single-photon generation.

\begin{acknowledgments}
We thank H.~P.~Specht for fruitful discussions and C.~J.~Villas-Boas and E.~Figueroa for their contributions to the theoretical model. This work was supported by the Deutsche Forschungsgemeinschaft (Research Unit 635), by the European Union (Collaborative Projects AQUTE and SIQS), and by the Bundesministerium für Bildung und Forschung via IKT 2020 (QK\_QuOReP).
\end{acknowledgments}

\end{document}